# Hydromechanical considerations on the evolution and diversification of the echinoderm bauplans

Michael Gudo[1]

Abstract

The pentaradial organisation of echinoderms is postulated to have evolved as the result of the reorganisation of the internal U-shaped mesentery of the intestinal tract during inflation of the trunk of a pterobranch-like ancestor. Under this scenario, loops of the mesentery developed between five hydraulic bulges by three different mechanisms: (1) by the early formation of two additional loops resulting directly in five hydraulic bulges, (2) by the subsequent formation of two additional loops resulting in the formation of initially three and then five hydraulic bulges or (3) by the inflation of the body without formation of loops. Accordingly there are at least three main evolutionary pathways within the echinoderms. The anatomical structures, such as the ambulacral system or the skeletal capsule, which characterize the echinoderms, developed independently in both lineages. From all the three pathways various body structures can be derived matching those found in fossil and extant echinoderms. The eleutherozoans and the earliest pentaradial echinoderms most likely evolved from the direct-pentaradial pathway whereas triradiate echinoderms and those which show a pentaradial organisation superimposed on triradial symmetry, likely evolved along the indirect pathway. The asymmetric or bilateral symmetric echinoderms evolved from the third paythway. The most important morphological transformations leading to the directly and indirectly pentaradial echinoderms are discussed and described.

## Introduction

It can be generally assumed that the fossil record of echinoderms does not show the evolutionary transformations by which this phylum evolved. Within the echinoderms the oldest known representatives were already diversified and show essential characteristics of this group, so that they shed no light on the affinities and origins of echinoderms, or on the manner in which their essential organization has been developed (Beaver et al. 1967a); even most recent findings of early echinoderms rise more questions than they give answers (Shu et al. 2004). Furthermore, among the earliest known representatives of the echinoderms there are those in which pentaradial symmetry was already very well developed, and even in the Lower Cambrian eocrinoids, brachioles are generally distributed in fives. This indicates that the fossil record offers no solution to the question of the origin of the pentamerous organization of the echinoderms (Beaver et al. 1967b; Hyman 1955). Therefore the origin of echinoderms as well as the origin of the pentaradial body organization needs to be reconstructed via an anagenetic scenario based on a historical evolutionary theory (i.e. a nomological-deductive or functional explanation, sensu Bock 1991; Bock 2000), discussing if the pentaradial organization evolved as a biomechanical necessity or if body symmetries are not related to any physical principles but evolved arbitrarily. Subsequently the biostratigraphy of fossil echinoderms can be interpreted in the context of the reconstructed historical narrative, and the given physical explanations.

Gil Cid et al. (2003) for example published a biostratigraphic pattern of echinoderms which shows that pentaradial eocrinoids were the earliest echinoderms (lower Cambrian), followed by Cincta and Cornuta (middle Cambrian), while Cystoids (Diploporita) do not occur until the Ordovician. This biostratigraphy indicates a biodiversification event in the Upper Ordovician; over a geologically short time almost all groups of echinoderms are represented by various genera. Of course these results only partly correspond to the global biostratigraphy of echinoderms and the question is still open, if the global diversification of echinoderms can be summarized in the sequence asymmetric eocrinoids, asymmetric homalozoans, triradiate echinoderms (such as the Helicoplacoidea) and finally pentaradiate echinoderms. Nevertheless even a raw quantification of taxon numbers might give some support but does not tell the complete story. Palaeontological findings as well as stratigraphic correlations or taxon-quantifications over time are not sufficient to explain the mechanisms how echinoderms evolved. In fact such results indicate that more than one main evolutionary pathway is present within the echinoderms, and that these pathways occur at particular times in earth history. These pathways have to be reconstructed on the basis of a structural-functional and hydraulic conceptualization of these organisms, to indicate the functional constraints and restraints for organic design and evolutionary transformations.

---

1) Dr. Michael Gudo, Morphisto – Evolutionsforschung und Anwendung GmbH (Institut für Evolutionswissenschaften), Senckenberganlage 25, 60325 Frankfurt am Main, Germany, e-mail: michael.gudo@morphisto.de

## Theoretical Background

Evolutionary research has to deal with two major aspects. At first an historical aspects which deals with the reconstruction of evolutionary pathways showing how particular body plans have been developed. This entails developing two concepts: (1) a concept of the organism, which constitutes the scientific subject-matter, and (2) a concept for reconstructing and evaluating evolutionary pathways. Subsequently evolutionary research can deal with the mechanisms by which these transformations were driven. In this resepct life has to be seen as a dynamic process of self-preservation, we call it a morphoprocess and accordingly evolution can be understood as a transformation of such morphoprocesses. Consequently the conventional way which starts with the mechanisms of evolution (such as adaptation, speciation etc.), has to be inverted: evolutionary research should start with the historical part (= anagenetic reconstruction) and with the concept of the organism. Mechanisms of evolution have to be investigated subsequently (for more details see Gudo 2004b; Gudo & Gutmann 2003).

A concept of the organism is a description of living beings with regard to a particular descriptive goal. For evolutionary research organisms should be described as operationally closed energy transducers and coherent hydraulic entities (Gutmann 1988, 1993). This means that organisms are bionomic units, which provide their structure with energy by themselves. They transform energy via their chemomechanical interactions on the molecular level powering a cascade of molecular, cellular and anatomical structures for maintainance of reproduction, shape, propulsion, and propagation (Gutmann 1991). This conceptualization of organisms is dedicated to a certain purpose which is to understand the functional design and to reconstruct the evolutionary history of animals. It might be compared to a description of cars or computers: They can either be described as status symbols or as furnitures, but to understand the purpose of their structure and their functional design, it is more suitable to describe them as energy transducing devices for locomotion (cars) or for complex calculations (computers).

Since evolution as a morphoprocess modifies the body structure of an organism, a structural functional analysis seems to be more suitable than an approach which reduces organisms to certain features or genes. Our particular methodology can be compared with reverse engineering, as known from Japanese industry. Reverse engineering involves analyzing a technical apparatus for a so-called rapid prototyping, so that finally this apparatus can be produced without any developmental work. Transferring this method to evolutionary investigations means analyzing organisms in a structural-functional sense, in order to understand the form-function complex, so that finally the

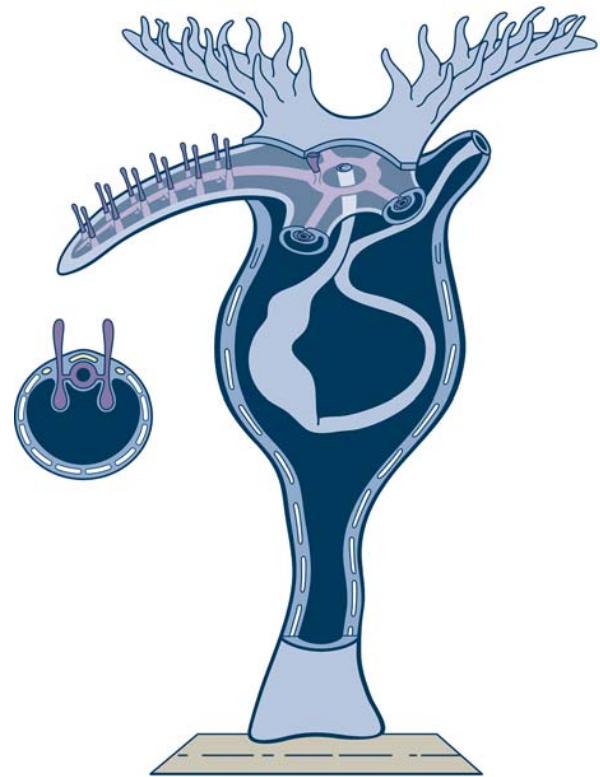

**Text-fig. 1.** Hydraulic conceptualization of the echinoderm body structure. Echinoderms have a body capsule in which the coelom fluid acts as a hydraulic skeleton (black). The body shape is generated and preserved by interaction of the hydraulic fluid filling with the mutable connective tissues, muscles (blue) and rigid skeletal elements (white). The muscles and the mutable connective tissues set the body fluid under pressure so that the arrangement of structures under tension determines the body shape and its locomotory capacities. Furthermore fibre networks and internal tensile chords constitute five hydraulic pneus, represented by the arms. Such pneus eventually arrange in a pentaradial pattern. Each of these arms carries a separate hydraulic system – the ambulacral system.

body organization is understood as it would be an engine-like structure. This requires the transfer of particular aspects of engineering-like examinations to biological objects (= engineering morphology, Gudo 2002; Gudo et al. 2002).

Under this approach echinoderms can be thought of as having a body capsule in which the coelomic fluid acts as a hydraulic skeleton and the body shape is generated and preser-ved by interaction of the hydraulic fluid filling with the so-called mutable connective tissues, thin muscles and rigid skeletal elements. The muscles and the mutable connective tissues set the body fluid under pressure so that the arrangement of structures under tension determines the body shape and its locomotory capacities (Gudo 2004a).

From this hydraulic conceptualization, the question arises, what is the origin of this particular body construction? To answer this question anagenetic reconstructions have to be done, which integrate particular biological (histological) and palaeontological subjects. Nevertheless, a convincing answer can only be given if the specific morphological transformations are reconstructed, if the evolutionary restraints are identified, and if the reconstructed pathways are unidirectional transformations which refer to the principles of economisation, optimisation and differentiation (Bonik et al. 1977; Peters & Gutmann 1971; Vogel 1983).

## Evolutionary lineages within the echinoderms

As a result from several molecular (e.g. Halanych 1995; Peterson et al. 2000) and structural-functional studies (e.g. Gudo 2004b; Gudo 2005a, b, in press) echinoderms evolved from a bilateral pterobranch-like ancestor which itself could be derived from early chordates via an enteropneust-like intermediate (Gudo & Grasshoff 2002; Gutmann & Bonik 1979). When the intestinal tract was bent into an U-shaped loop the mesentery followed and widely determined this course so that finally two parallel planes of the mesentery were present.

The body shape and positioning of the gut were established, when the body was more and more inflated by the internal fluid pressure. The body was enlarged in its anterior part, whereas the hind part narrowed and was eventually transformed into a stalk and holdfast. The internal space of the anterior part became wider and shorter, as compared to the elongated shape of the ancestor. Accordingly, the intestinal tract, should it not have been dramatically shortened, had to be laid in loops. The number of loops which the intestinal tract developed is of importance for the symmetric organisation of the resulting body structures, because the mesentery follows the position of the gut running along the body wall. The collagenous fibers of the mesentery have the mechanical effect of tensile chords holding the body wall in its place. By contrast, the intermediate parts of the wall between these zones of tensile chords can bulge out under the internal hydraulic pressure. These lateral bulges may grow, leading to the evolution of a radiate body organisation, departing more and more from the originally elongated bilateral-symmetric organisation of the pterobranch-like ancestor. We can summarize that the bilateral organisation was overcome by developing outgrowths perpendicular to the anterior/posterior body axis in a radial – more precisely – in either a tri-radiate or in a penta-radiate pattern (Figure 2).

Obviously, both of these paythways have been passed and both bauplans influenced the evolutionary history of echinoderms. The fossil record provides a large amount of fossils which can be assigned to either of these bauplans or evolutionary pathways. Nevertheless during their evolutionary history, both of these bauplans lead to pentaradial echinoderms, so that two evolutionary scenarios have to be discussed: (1) The pentaradial organization developed directly when five loops developed directly from the original U-shaped gut (Gudo 2004b). (2) Five loops of the intestinal tract developed subsequently (indirectly) – first three, then five. Besides these two pathways a bilateral inflation of the trunk was also possible and lead to the third pathway which is the lineage of homalozoans in a broader sense; however this option will be not discussed here (for more details see Gudo 2005a, b, in press).

The fossil echinoderms giving rise to the suspicion that primarily triradiate forms were secondarily overgrown by a pentaradial organisation are represented by cystoids, such as *Glyptocystites* or *Glyptosphaerites*. If we consider the scenario for the direct-pentaradial evolution, it is easy to assume that the loops of the intestinal tract appeared subsequently and that initially only three loops developed while the body was inflated anteriorly. From these three loops tensile chords would have reached to the body wall in three regions leaving tensile chords absent from the three intervening regions. In this way a triradiate body organisation could have been attained in which tentacles of the collar were mechanically supported from the hydraulic bulges, so that they finally developed into three ambulacral fields. From this body organization, which might have had a largely voluminous and stout shape, certain triradiate echinoderms such as the Helicoplacoids can be derived.

The indirect-pentaradial echinoderms could now evolve from an intermediate stage in which the body capsule was not yet completely stiffened by skeletal elements. In one of the three areas where tensile chords were absent the existing loop of the gut forms another loop, so that eventually five loops resulted and the mesenterial tensile chords met the body wall in five regions. During further body growth (individually as well as evolutionary) the body structure attains a pentaradial symmetry which superimposed over an initial triradiate symmetry.

For the original ambulacral fields there are two options of further development: it can divide into two, each branch running at the sides of the loop. Or it can continue to run down in the middle of the original segment, and branches above the upper curving, sinking right and left of the loop one branch down. Indeed, fossil echinoderms with four ambulacral fields exist, such as the cystoid *Lovenicystis*. Moreover, fossil echinoderms with five ambulacral fields exist, obviously derived from triradial ones: one of the three ambulacral fields (which are starting from the ring-canal) divides into three (*Glyptocystis* or *Glyptosphaerites*). These forms do not show the strictly regular pentaradial organisation of Eleutherozoa. The body structure which resulted from these transformations is similar to the body organisation that resulted from the direct pathway envisaged. However, particular differences can be expected, so that echinoderms may be placed into two main evolutionary fields.

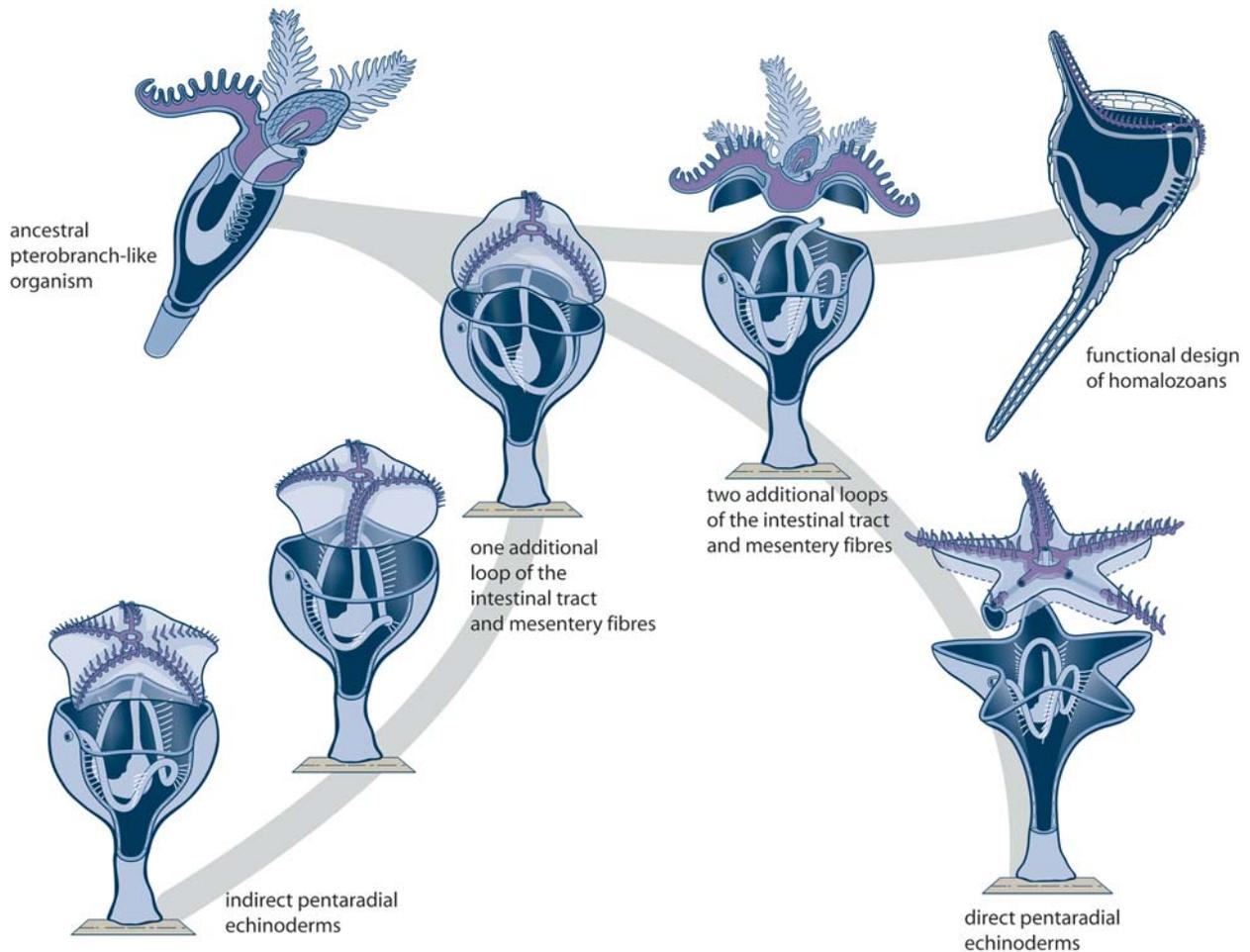

**Text-fig. 2.** Main evolutionary pathways of echinoderms. The left lineage shows the indirect-pentaradial echinoderms in which the intestinal tract and mesentery formed one additional loop first and the second loop subsequently. Accordingly the body was inflated more or less spherical and the ambulacral fields subsequently branched off each other until they finally formed a pentaradial pattern running over the body surface. The middle lineage shows the direct-pentaradial echinoderms in which the intestinal tract and mesentery formed directly two additional loops. Accordingly the body was only the untethered reagions from the beginning on and therefore strict pentaradial body structures evolved. The third pathway is only shown for completeness. The homalozoans in a broader sense comprise all those echinoderms which do not show a tri- or penta-radial body symmetry; they are dedicated to a simple inflation of the body by expanding the distance between anus and mouth of U-shaped intestinal tract.

## Some consideration on the diversification of echinoderm bauplans

The diversification basic bauplans of echinoderms, has to be reconstructed in the same manner as presented before. This means that aspects of functioning, functional design (body structure), and the physical principles of hydromechanics have to be taken into account. Furtheron anatomical histological and palaeontological results have to be interpreted in the frame of plausible transitional (anagenetical) scenarios. In the following I will not present a full scenario for all echinoderms bauplans. The goal of the following passages is to summarize several aspects which are of crucial importance for the reconstruction of such a scenario. The focus here will be set on the direct-pentaradial echinoderms which are also known from the Recent.

### Crinozoans

The functional design of the crinozoans is characterized by extraordinary enlarged skeletal elements in the posterior part of the body and in the arms. In an evolutionary scenario it has to explained how the coelomic cavities were reduced and how the course of the intestinal tract in the body was simplified by forming a spiral with only one twist. Due to particular histological differences between recent crinozoans and recent eleutherozoans, it is most likely that crinozoans represent an evolutionary lineage which is convergently to the lineage leading to the Recent eleutherozoans. The small body cavity makes it necessary that the gonads find their place in bursae between the arms; the scenario has to consider a basic difference in the position of gonads in crinozoans and eleutherozoans.

The huge fossil record of crinozoans indicates that various evolutionary modifications were possible giving rise to sea lilies with more or less branched arms, with long or short stalks and with different numbers and types of cirrae. However, in the pathway leading to isocrinids (which are also represented by some Recent specimens) the morphology of the ancestral crinozoan with reduced coelom cavities and enlarged skeletal elements is assumed to be preserved. Under evolutionary aspects the crinozoans with a reduced stalk, such as the comatulids, represented for example by *Antedon* sp. are of particular interest. They show how sessile organisms could successively attain a temporarily pelagic or vagile life style. During their juvenile phase comatulids are anchored to the substrate by a holdfast and a relatively short stalk, but later, when they have developed into an adult, they can lose their connection to their holdfast and start swimming by rhythmic contraction of their arms. Nevertheless, they do not swim over long distances or for long periods, but use this capacity for taking flight or for changing their feeding position. The comatulids represent one route by which the sessile life style was abandoned, they can therefore be seen as a kind of a model system for such a transition of a sessile organism into a vagile organism.

## Ophiuroids

Generally the ophiuroids are taken as a side branch of asterozoans (Dean 1999), sometimes they are phylogenetically related to echinoids (Littlewood et al. 1997; Wada & Satoh 1994). Nevertheless, these results are phylogenetic results and do not represent or explain evolutionary (anagenetic) transitions since relations to other Recent or fossil specimens are discussed only on the basis of sortings. In contrast to these phylogenetic suggestions, histological investigations and structural-functional considerations make it more suitable to understand the ophiuroids as descendants from a crinozoan-like ancestor that – of course – differs from the ancestor of Recent Crinozoans in particular histological and anatomical details (such as the position of the nerve chords, coelomic canals and the position of ambulacral system). Ophiuroida are not assumed to be genealogically related to the Crinozoa, but under structural-functional aspects it is more plausible that the functional design of ophiuroids evolved at the evolutionary stage of a crinozoan-like ancestor and not at the stage of an almost complete asterozoan.

Nevertheless, the fossil record of asteroids and ophiuroids shows many intermediates where for example ophiuroids have a voluminous coleom running in their arms (Dean 1999). According to these findings the question arises what are the constructional limitations and restrictions – and, is there probably a second possibility for the ophiuroids functional design to evolve? The significant difference between ophiuroids and asteroids has to be seen in the movability of the arms and the position of nerve chords and coelomic canals. The functional point is that no mechanical advantage can be seen if the skeletal elements of an asterozoan are enlarged more and more until the complete arm is filled with vertebrae-like elements. According to this it would be necessary to remove the gut and the gonads from the arm and to place them in the bursae. But this means that simultaneously when skeletal growth is increased in the arms, the skeletal stiffening of the body disk has to be reduced. The skeletal elements (=vertebrae) are solitary structures which are connected via joints by connective tissues and muscles. Biomechanically it is easier to explain these structures as originated quite early from a crinozoan-like ancestor where the arms need to have a high movability for filtering nutritive material. By the way: ophiuroids are the only recent echinoderms (beside the crinoids) which are to a large extent filter-feeders (either feeding from the water columns or collecting nutrients from the sediment-surface. Asterozoans and echinoids are predators or graspers, holothurians are sediment feeders.

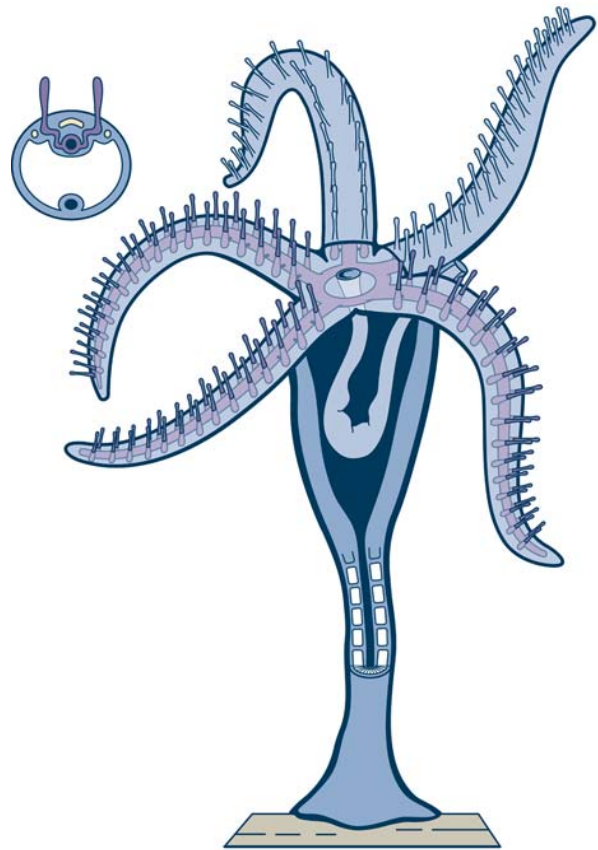

**Text-fig. 3.** Body structure of an ancestral crinozoan (left) and the suggested ophiuroid ancestor (right). While the skeletal elements bevome larger the movability of the arms and the stalk increasing providing a better feeding performance. In the ophiuroid-ancestor the anus was reduced early while in the crinozoan-ancestor it was preserved. – Illustration by Antje Siebel- Stelzner.

## Eleutherozoans

The evolutionary field of the eleutherozoans comprises the Echinozoa (echinoids and holothuroids) and the Asterozoa. They are all characterized by a vagile benthic life style. However, within holothurians some remarkable body structures evolved, which allowed them to adopt peristaltic burrowing in the

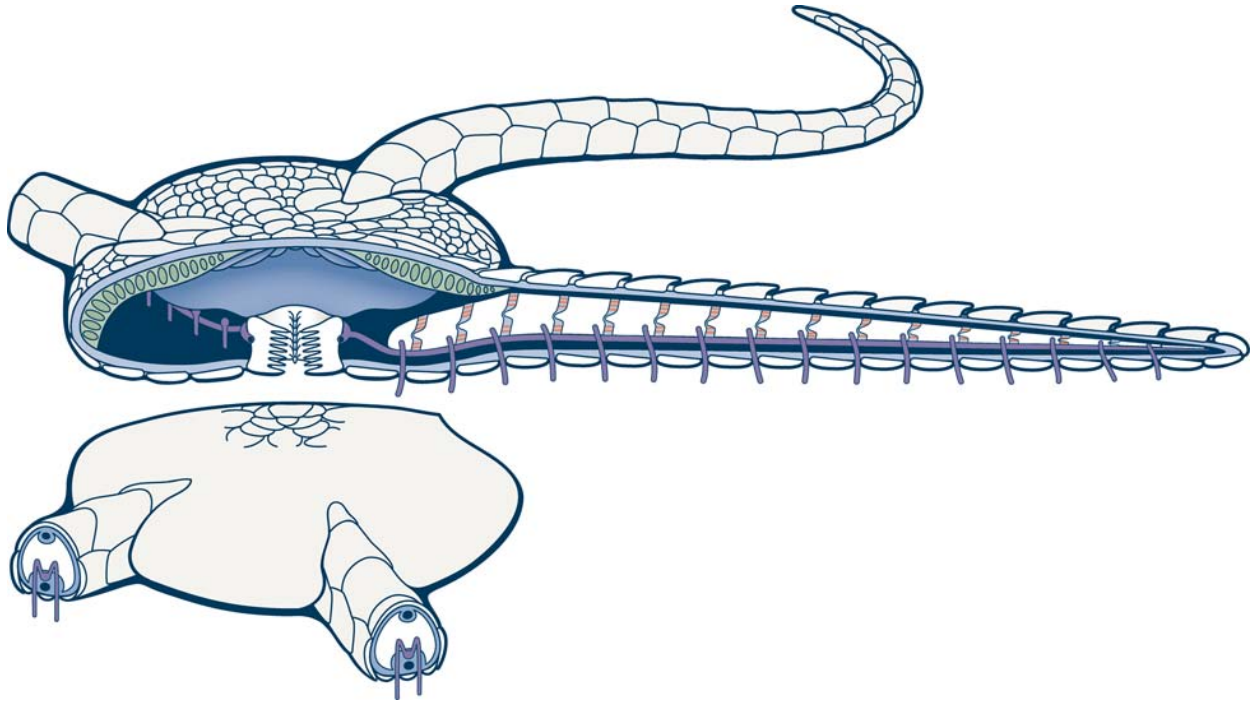

**Text-fig. 4.** Functional design of an ophiuroid. Skeletal elements of the arms are connected via muscles and in summary they provide a high movability for the arms. Gonads are placed in the bursae between the arms; the intestinal tract has no anus and therefore faeces have to be digested through the mouth from time to time.

sediment and undulatory swimming movements in a quasi-pelagic life style. The crucial point for the origin of the eleutherozoans is a functional transformation of ambulacral feet which means that they no longer function as filtering structures but as locomotory system. In many eleutherozoans the ambulacral pinnules developed suckers which improved locomotion in contoured environments such as reefs and thereby opened up a new environmental ressource. This transformation is based on some simple modifications of the arrangement of the muscles in the originally haunch-like ends of the podia. Ambulacral feet were initially used to feed from the substrate by bending the stalk so that the oral field was faced to the substrate. Nutrients were collected by the ambulacral podia, and even larger prey could be caught and held. In the next step the stalk lost its anchoring to the substrate and the ambulacral podia were used for locomotion on the substrate. In contrast to the ophiuroids the arms of the eleutherozoans are still voluminous and filled with coelomic fluid or gonads and diverticles of the intestinal tract, respectively. Therefore eleutherozoans do not brush the substrate like early ophiuroids, but they collected and caught nutritive material using their ambulacral feet in a more selective manner. Later, new resources for feeding opened up, because larger prey could be caught and held. This new way of feeding gave rise to a number of morphological innovations.

The evolutionary pathway of eleutherozoans is envisaged as having branched into two main lineages, the asterozoan lineage and the echinozoan lineage. In the asterozoan lineage the arms of the ancestral pelmatozoan Ur-Echinoderm were preserved. These arms are thick, they carry the ambulacral system and they contain gonads and diverticles of the intestinal tract. Since these diverticles are attached to the aboral body wall by a double mesentery, they can easily be derived from the original loop which gave rise to the pentaradial echinoderms. The gonads where placed in the arms. The fluid filling of the arms, and therefore also the diverticles of the gut and the gonads worked as a hydroskeleton for the muscles and mutable connective tissues in the arms. The body attained a lens-like shape from which the arms grow outwards. Most of the Recent asterozoans have an anus (except the Astropectinidae) at their aboral side, while there is little firm evidence of an anus in Palaeozoic asteroids. This makes it most likely that the anus was initially completely lost when the stalk was reduced and that the anus of recent asterozoans is a secondarily evolved structure.

The evolutionary pathway of asterozoans opened up a wide range of potential morphologies. It is of interest for the understanding of the echinozoan pathway that some asterozoans have spines or spine-like structures, similar to those of the echinoids. Furthermore there are asterozoans, such as the cushion sea stars, which have lost the narrow regions between their arms, so that they attain the shape of an inflated cushion. This shows that secondary inflations of an asterozoan-like functional are generally possible.

The echinozoans are suggested to be evolved as a separate branch from early eleutherozoans that had lost their anchoring to the substrate. Whereas in the asterozoan lineage the trunk was reduced, in the echinozoan lineage the trunk was inflated due to a relative growth of aboral and oral plated surfaces (as demonstrated e.g. by the asteroid *Spenaster*) and resulted in spherical body shapes. The ambulacral field become maintained as integral part of the body wall. In this respect the results of the EAT are quite useful, because they show which skeletal parts are involved in these transformations.

Since the spherical shape is the most economic shape for a hydraulic body structure, the pressure which was continuously generated by the mutable connective tissues and the muscles lead to the formation of large skeletal elements that finally fused to a rigid skeletal capsule in which only small sutures provide a small amount of space between the ossicles so that individual growth was possible (Johnson et al. 2002). From the skeletal plates of the capsule spines developed. Originally the spines were structures for generating tensile strength within tissues (Gudo 2004a), but during the transformations mentioned above these spines became involved in locomotory activities, since most of the ambulacral podia were on surface regions which were not in contact with the substrate.

The spherical shape of the sea urchins is generated by internal osmotic fluid pressure (Dafni 1980, 1984, 1986, 1988; Dafni & Erez 1982). However, at the oral side where the mouth opening is located, a jaw apparatus developed from skeletal elements of the radialia and interradialia reaching into the coelomic cavity. This internal jaw apparatus most likely developed from skeletal elements of the dermal tissues and their muscles determine the flattening of the oral side; the primitive jaw apparatus of the ophiocistioids can be seen as an intermediate stage. Most likely internal tensile chords provided by the mesentery of the intestinal tract supported the oral flattening. Additionally the growth of skeletal elements of the radialia and interradialia and the use of spines on these plates for grasping prey initialized the development of a particular jaw apparatus. An important precondition for the evolution of the lantern of Aristotle might be the existence of ambulacral podia with their sucking capability. Grasping nutrients from the ground with an apparatus such as the lantern of Aristotle is only possible if the sea urchin is able to hold itself on the substrate with ambulacral podia. The evolution of the lantern of Aristotle can therefore only be understood in relation to the function of the ambulacral podia. This evolutionary innovation can – in some respect – be compared with the origin of the radula of molluscs: both of these structures are capable for grazing material such as biofilms, corals, sponges etc. from the substrate, but both could develop only if active anchoring to the substrate was possible (in molluscs via the creeping foot, in echinoids via the ambulacral feet).

The evolutionary field of echinoids comprises a large number of varieties. They vary predominantly in the size of their capsule and spines. Some of them have short spines (such as *Sphaerechinus* or *Echinus*), some have only few thick spines (such as the Cidaridae), some have long and slender spines (such as the Diademidae) and even some which have slender spines are lacking a skeletal capsule such as the leather sea urchins (*Asthenosoma varium*).

With regard to inflation of the body capsule until it attains a spherical shape, the intestinal tract could take more room inside the coelomic cavity. Therefore it attains a remarkable course inside the inflated body: The already existing loops were duplicated by further elongation of the intestinal tract. This finally leads to the double layered looping course of the intestinal tract as shown for *Echinus esculentus* or *Sphaerechinus granulosus* (Strenger 1973).

The origin of the holothurians has to be seen in early echinoids which have not yet developed a fully sclerotized body capsule. As Haude (2002) mentioned, there are two subsequent structural alternatives for an early echinozoan body structure: The first is the formation of a solid body wall, i.e. a skeletal capsule without any possibilities for active shape deformations – this is the pathway to the echinoids. The second is a facultative vaulting of the body wall by activity of muscles and mutable connective tissues. Facultative vaulting of the body shape lead to elongated shapes if mutable tissues are capable to limit diametrical enlargement, so that the body is inflated in the oral-aboral direction predominantly. However, such body structures were reorientated; they no longer had their mouth opening facing the substrate, but laid on their elongated side. A worm-like body shape has developed. Haude (1993; 1995; 2002) has mentioned several anatomical and skeletal modifications which must have taken place during this morphological transformation. The most crucial result of this morphological transformation can be seen in the nearly complete reduction of the skeletal elements due to continuous movements of the body wall during peristaltic contractions. Eventually only tiny spiculae in the shape of anchors, wheels or spines remained. Intermediate stages which still show larger skeletal plates are represented by *Procrustia* and by *Podilepithuria* (Haude 2002). Simultaneous with the reduction of the skeletal elements, the amount of mutable connective tissues in the body wall dramatically increased and the circular muscles widely reduced. Longitudinal muscles are present in five regions corresponding to the original ambulacral fields. The sea cucumbers locomote by peristaltic movements and in some extreme cases even undulative swimming and peristaltic burrowing in the sediment becomes possible (Miller & Pawson 1990).

# Discussion

The evolutionary history of echinoderms has to be reconstructed by showing the transformation of one functional design into another one. This has been done here for the early differentiation into two pathways which both lead to pentaradial echinoderms. More detailed scenarios which consider the various functional designs of the Recent and fossil echinoderms have to reconstructed subsequently. Nevertheless before an evolutionary scenario can be reconstructed, particular anatomical, histological details and aspects of functioning and function have to be considered. This was the goal of the presentation given here. Some result are foreseeable from such considerations, as for example that the traditional assumptions of the anagenetic position of ophiuroids has to be reconsidered. The functional design of ophiuroids is suggested to have to be derived from a crinozoan-like ancestor. This does not mean that Ophiuroidea are a subgroup of Crinoidea. The scenario in

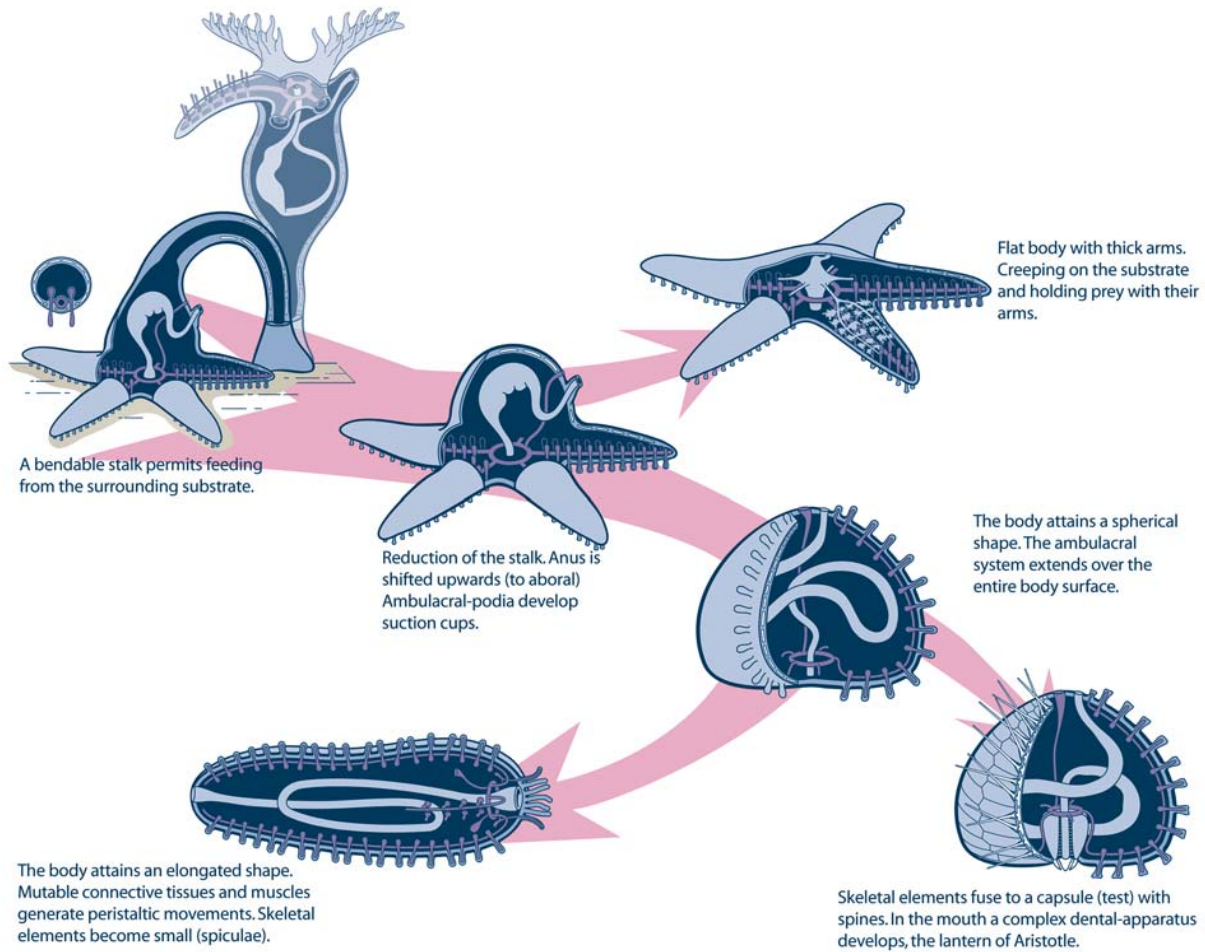

**Text-fig. 5.** Evolutionary field of eleutherozoan echinoderms. The functional designs of asteroids, echinoids and holothurians can be derived from one common pelmatozoan ancestor. Initially the pelmatozoan ancestor starts feeding from the substrate by bending its body. Accordingly the ambulacral podia developed a sucker so that finally the body could either detach from the trunk or the trunk reduced completely. This transformation is not completely solved yet, however, it might be possible that both options were possible.

which these organisms are derived from a crinozoan-like ancestor means that the ancestors of both of these groups were similarly structured. Due to functional aspects it is difficult to reconstruct an evolutionary pathway for ophiuroids from asterozoans, because there are no economic transformations from the large fluid filled arms of an asterozoan with contain gonads and diverticles of the intestinal tract to the thin and highly movable arms of an ophiuroid. This differentiation must have taken place much earlier, which means shortly after the pentaradial organisation was established. There were two options possible, one is that the arms which evolved as hydraulic outgrowths of the body wall maintain the lumen so that gonads and gut-diverticles were placed in the arms; this opened the pathway of asterozoans and echinozoans. The second option is that the arm lumen becomes reduced so that the arms themselves attain a high mobility and develop branches, brachioles and pinnules. This leads to the conclusion that ophiuroids can be derived from the same ancestor from an ancestor which is – in some aspects – more similar to the ancestor from which the crinozoans evolved, while the asterozoans evolved in a separate branch in which the arms were hollowed out at an early stage. In the crinozoan branch a filtering apparatus developed and therefore the ambulacral system was not completely enclosed into the tissues of the arms. In the ophiuroid branch initially feeding of larger particles was performed. This also allowed feeding from the substrate finally leading to the dissolution of the body with the arms from the trunk. If such an anagenetic relation is accepted, this would also explain that sometimes the ophiuroids plot closer to the asterozoans and sometimes closer to echinozoans (Littlewood, et al. 1997; Wada & Satoh 1994).

If evolutionary scenarios in some aspects contradict the traditional assumptions this does not mean that traditional results are wrong; however, they are valid in another context. Anagenetic reconstructions show the transitions from one bauplan to another, phylogenetic relations show genealogic relationships, but not more. The problem is that evolutionary transformations cannot be read from the fossil record or from sor-tings on the basis of molecular analyses; these results only provide some information which can be integrated into an reconstructed anagenetic scenario. Evolutionary scenarios have to be reconstructed with regard to constraints provided by the functional design of the organisms and with regard to the evolutionary restraints for functional transformations. The hydraulic organisation of organisms, in particular of the echinoderms, makes it necessary that anagenetic transformations are reconstructed by pointing on the functional and anatomical changes.

This first structural-functional conceptualization of the echinoderm body opens up a new field of evolutionary research. Further refinements need to be done so that evolutionary transitions for the evolutionary diversiications with in the evolutionary field of echinoderms can be reconstructed.

## Acknowledgements

I thank the DFG for funding the project GU566/1-1. For many helpful comments I thank Raimund Haude (Göttingen), Manfred Grasshoff (Frankfurt), Stefan Peters (Frankfurt) and Tareq Syed (Frankfurt). I thank Andrew Smith (London) for improving the language and several helpful comments, and Antje Siebel-Stelzner for the illustrations.